# Communicating the Heisenberg uncertainty relations: Niels Bohr, Complementarity and the Einstein-Rupp experiments

Jeroen van Dongen[1]

**Abstract**: The Einstein-Rupp experiments have been unduly neglected in the history of quantum mechanics. While this is to be explained by the fact that Emil Rupp was later exposed as a fraud and had fabricated the results, it is not justified, due to the importance attached to the experiments at the time. This paper discusses Rupp's fraud, the relation between Albert Einstein and Rupp, and the Einstein-Rupp experiments, and argues that these experiments were an influence on Niels Bohr's development of complementarity and Werner Heisenberg's formulation of the uncertainty relations.

"Dear Einstein", Niels Bohr's letter of 27 April 1927 begins, "[b]efore his holiday trip to the Bavarian mountains, Heisenberg asked me to send you a copy of the proofs that he was expecting, which he hoped might interest you." Bohr sent Albert Einstein the proofs of Heisenberg's article on the uncertainty relations. He sent these not just because Heisenberg had asked him to do so: the article, in Bohr's opinion, was "closely related to the questions that I have had the great pleasure of discussing with you a number of times." Bohr and Einstein had discussed the physics of quanta and its problems since they had first met in April of 1920. Bohr now introduced Heisenberg's newest work by placing it in the context of a very recent contribution by Einstein. Indeed, Bohr wrote that he wished to avail himself of "the opportunity to include some remarks concerning the problem that you discussed recently in the proceedings of the Berlin Academy."[2]

Bohr told Einstein that he believed that the uncertainty relations made it possible "to avoid the paradox discussed by you, concerning the spectral resolution of the light emitted by a moving atom and observed through a slit perpendicular to the direction of motion." Here Bohr referred to the situation depicted in Figure 1: an atom moves behind a slit and emits light in its direction, and the "paradox" he believed that Einstein had introduced concerned the light that eventually emerged. Bohr's rendition of this paradox stated that the wave theory entailed, due to "the limitation in the time of observation" (or, the fact that the wave has a finite length), an "uncertainty" in the frequency $v$ of magnitude $\Delta v = v/a$, with $v$ the atom's velocity and $a$ the slit's width. Taking account of diffraction at the slit and the Doppler effect due to the source's motion would produce the same relation, Bohr pointed out. Then, one may fear a contradiction with strict validity of energy conservation, as the emission's 'frequency' should be narrowly prescribed by the relation $E = h\nu$. However, Bohr wrote, in the particle picture suggested by the energy relation, it is not too hard to imagine that the emitting atom can undergo a radiative recoil "that may deviate from the perpendicular direction of observation." This would introduce variations in the energy of the light quantum, and hence, its frequency. Consequently, Bohr found, there need not be a contradiction, but rather just two different ways of viewing the situation, which both are in accordance with, and thus formally sanctioned by, Heisenberg's relations. Bohr referred to a footnote in Einstein's paper that he took to be further support for his rendering of the problem at hand: Einstein had said, according to Bohr, that "no possible 'light quantum' description can ever explicitly do justice to the geometrical relations of the 'ray path'."[3]

---

[2] Niels Bohr to Albert Einstein, 27 April 1927, pp. 21-24 in Kalckar (1985), on p. 21. The proofs mentioned are of the article Heisenberg (1927).

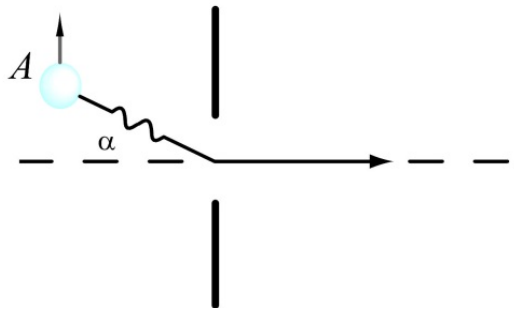

Figure 1. Atom *A* radiating behind a slit, with the emitted light undergoing diffraction.

Attentive readers will have identified in the above not just Bohr's introduction of the uncertainty relations to Einstein, but also early intimations of his ideas of complementarity. Indeed, Jørgen Kalckar, the editor of the beautiful Volume 6 of Bohr's *Collected Works*, saw in this letter already "the essence of the complementarity argument." He points out that Bohr had formed his earliest ideas on complementarity during a vacation in Norway, away from Heisenberg in Copenhagen, while the latter was producing the uncertainty relations.[4] Heisenberg's relations, Bohr's letter to Einstein suggests, gave further theoretical authorization to these early complementarity ideas. As Bohr put it in his letter: "Through the new formulation we are presented with the possibility of bringing the requirement of energy conservation into harmony with the consequences of the wave theory of light, since according to the character of the description, the different aspects of the problem never occur at the same time."

So, according to Bohr, his and Heisenberg's ideas were related to earlier work by Einstein, particularly to a light quantum puzzle that Einstein had published in the proceedings of the Berlin Academy. In fact, Bohr even repeated some of Einstein's own words to underscore the point of his analysis. Let us look at what Einstein himself wrote in the footnote that Bohr referred to: "In particular, one may not assume that in the quantum process of emission, that energetically is determined by location, time, direction, and energy, is also in its *geometrical* properties determined by these quantities".[5] Indeed the note suggests that the wave-like behaviour (i.e. the 'geometrical' properties) of emitted radiation cannot be captured in a particle picture perspective (i.e. the 'energy' perspective). This, in turn, does seem a natural stepping stone to Bohrian complementarity ideas.[6]

---

[3] Niels Bohr to Albert Einstein, 27 April 1927, pp. 21-24 in Kalckar (1985), pp. 22-23.
[4] "Introduction", pp. 7-51 in Kalckar (1985), on p. 21; see also its p. 16 for a similar comment.

Yet, what did Einstein actually intend to express when he wrote the above words? What is their relevant context, and what exactly was the physical problem that he was concerned with? And how exactly was that relevant to Bohr? Einstein's light quantum had received a substantial boost due to the Compton experiments, yet many of its aspects had remained unclear. Bohr of course had strongly disliked the light quantum earlier, but he more or less had been forced to accept the idea after the experimental dismissal of the 'BKS'-theory, in which Bohr, Hendrik Kramers and John Slater had suggested that the description of light emission might still only need waves if only one was willing to give up on energy conservation.[7]

The "paradox" that Bohr referred to in his letter was actually based on one of two particular experiments that were to probe the wave versus particle nature of light emission. The experiments had been proposed in 1926 by Einstein, and Emil Rupp, in close consultation with him, had published the results of their execution. They were subsequently known as the "Einstein-Rupp" experiments.[8] These experiments are virtually unknown today. Indeed, many readers will ask themselves: "The Einstein-Rupp experiments?? Why have I never heard of these?" In this article, I will aim at two things: first I will give a brief introduction to the Einstein-Rupp experiments and their peculiar history. Secondly, I will return to their discussion by Bohr and thus exhibit their constitutive role in the history of quantum theory. The conclusion probes why these experiments have remained so obscure, despite the extensive historiography on Bohr, Einstein, and the quantum revolution.

**The Einstein-Rupp experiments**

In the fall of 1926, Albert Einstein published the outline of two experiments in the Proceedings of the Berlin Academy.[9] They addressed one of the most urgent questions in physics at the time: the experiments were to show if the emission of light was a process that was extended in time, or if instead light emission occurred in an instantaneous act. Of course, the first possibility would confirm a traditional oscillator-and-wave-like view, whereas the second possibility would cohere well with Einstein's own ideas on light quanta.

---

[5] Einstein (1926b), p. 337.
[6] For statements of complementarity soon after Bohr's letter to Einstein, see the drafts for his Como lecture, found on pp. 57-98 in Kalckar (1985); in particular passages on pp. 76, 79 exhibit a relation to the content of Bohr's letter. The published version is Bohr (1928; pp. 109-146 in Kalckar 1985); passages on pp. 567 and 570-571 (pp. 115 and 118-119 in Kalckar 1985) again reveal a conceptual link to Bohr's letter when they discuss the apparent conflict between limitedly extended wave fields, and the validity of the conservation laws and well-defined space-time coordination of observations.
[7] On Bohr's opinion of light quanta and its relation to the BKS-theory, see Kragh (2012), pp. 325-337.
[8] For this name, see e.g. Heisenberg (1930), p. 59.
[9] Einstein (1926b).

It is quite surprising that these experiments are so unfamiliar today. Apart from addressing a central question and being proposed by no lesser figure than Einstein, they also circulated at a crucial moment in the history of quantum theory. Still, the experiments are not mentioned in any of the standard Einstein biographies[10] and there is no substantial treatment of them in histories of the quantum theory (for example, the six weighty volumes on *The Historical Development of Quantum Theory* by Jagdish Mehra and Helmut Rechenberg (2000) discuss these experiments in less than two paragraphs;[11] Mara Beller's *Quantum Dialogue* (1999), has no mention of them at all).

The likely cause for this lack of attention is at least as surprising: the experiments were—supposedly—conducted by Emil Rupp, yet a decade later Rupp was exposed as a scientific fraudster; the results, obtained by Rupp in close consultation with Einstein and published back to back with the latter's theoretical paper,[12] were in the end generally believed to have been fabrications. The events led the German Physical Society to issue a statement in 1935 that Rupp could no longer publish in its journals, and that citations to his work were to be avoided;[13] Rupp was expelled from the professional community, and his work, even his work conducted under Einstein's auspices, gradually faded to the margins of anecdote and the silences of embarrassment. Recently, however, historical scholarship has looked at Rupp's career,[14] and has reconstructed his collaboration with Einstein in detail;[15] this work has confirmed the judgement that he made his results up. Yet, despite these studies, it is still the case that the experiments have largely but unduly remained outside the purview of the physics community, and are not given enough weight in the historiography of quantum theory and the foundational debates that surrounded it.

As I will argue here, these experiments played a substantial role in developments in 1926. Most importantly, they confirmed a wave picture of light, when many, including Einstein himself, initially expected a particle-like, instantaneous picture of light emission to be confirmed.[16] After all, only a few years before Compton scattering had been shown, and as little as a year before the Einstein-Rupp experiments Walther Bothe and Hans Geiger had done the experiments that dismissed the BKS theory.[17] But the experiments of Einstein and Rupp also influenced events in other ways. For instance, their initial interpretation was most likely of direct importance for Max Born, when he proposed the probabilistic interpretation of the wave function.[18] The experiments further played a role in the thinking of Werner Heisenberg, as he formulated his uncertainty relations; as we will see when we return to Heisenberg later. Clearly, these experiments deserve renewed attention, and their current obscure status is not warranted by their historical importance. This is further emphasized by the way that Bohr communicated the uncertainty relations to Einstein, and I will unpack Bohr's letter in what follows. First, I will briefly introduce Rupp, his fraud and his relation with Einstein. Then I will discuss the experiments as proposed by Einstein, and show how these were to assess the duration of light emission processes. Subsequently, I will present the various interpretations of the experiments, beginning with Einstein's. I conclude by offering some historiographical morals.

---

[10] See, e.g., Pais (1982a), Fölsing (1993), Isaacson (2007).

[11] See Mehra and Rechenberg (2000), pp. 235-236.

[12] The reprints of Einstein's and Rupp's articles, Einstein (1926b) and Rupp (1926b), circulated in a small booklet that contained both papers, which shared its cover page. A copy can be found in the Nachlass of Walther Gerlach at the archive of the Deutsches Museum, Munich.

[13] Document in the Nachlass of Walther Gerlach at the archive of the Deutsches Museum, Munich, entry 124-01; on it is written in Gerlach's hand "Mitt[eilung] der Geschäftsversammlung der DPG 1935 an ihre Mitgl[ieder]."

[14] French (1999).

[15] van Dongen (2007a, b).

[16] For Einstein's initial expectations, see Einstein (1926a).

[17] Bothe and Geiger (1925); Fick and Kant (2009).

[18] See van Dongen (2007b), pp. 126-127.

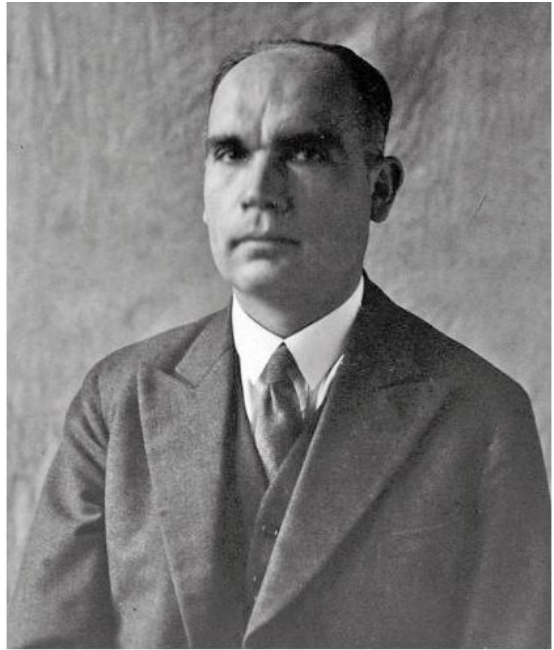

Figure 2. Emil Rupp (1898-1979). *Source*: University Library, Göttingen.

**Rupp, his fraud, and his relation with Einstein**

"Rupp, in the late twenties, early thirties, was regarded as *the* most important and most competent physicist. He did incredible things. … Later, it turned out that *everything* that he had ever published, everything, was forged. This had gone on for ten years, ten years!"[19] As this quote of Walther Gerlach (of Stern-Gerlach fame) suggests, Emil Rupp's rise and subsequent fall was quite visible to the contemporary physics community—easily comparable to the case of Hendrik Schön in our times. Indeed, after producing contentious work for close to a decade, the house of cards that Rupp had been trying to balance came to a dramatic collapse in 1935. Yet, it had all begun so very promisingly.

Emil Rupp (Figure 2) was born in 1898 in Reihen, a small community between the German towns of Heilbronn and Heidelberg. In Heidelberg, under the guidance of Nobelist Philipp Lenard, Rupp graduated for his PhD and fulfilled the requirements for his *Habilitation*, the German qualification that bestows upon its recipient the right to teach at a university. Rupp had published remarkable observations of the coherence length of light to obtain this qualification.[20] In his experiment, he had used a canal ray light source (i.e. a beam of ions that emits light after moving through a hole in a cathode—the "canal") that radiated light into a Michelson interferometer (Figure 3). By moving one of the mirrors of the interferometer until he no longer saw an interference pattern, Rupp would establish the light's maximum coherence length: it was given by the path difference that the light picked up in the interferometer. His results were quite impressive: in the case of hydrogen canal rays (at the $H_\beta$-line in the Balmer spectrum), he found a maximum coherence length of 15.2 cm, and in the case of mercury (for $\lambda = 5461 \, Å$), the value was 62 cm. This last value was about the same as had been achieved with sources at rest; the value for hydrogen was even an order of magnitude better.

---

[19] Walther Gerlach, interview with Thomas Kuhn, 18 February 1963, Archive for History of Quantum Physics.

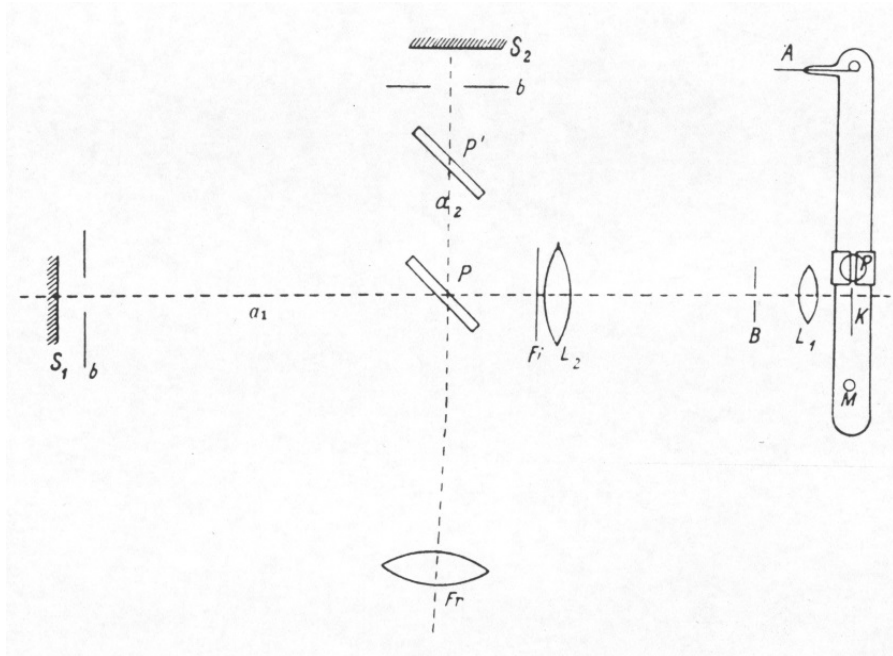

Figure 3. Rupp's original experiment for his *Habilitation*. Light enters the interferometer from the right, where it is emitted by canal ray light source at *K*. Source: Rupp 1926a.

Upon learning of these results, Einstein contacted Rupp on 20 March 1926. He realized that since Rupp had such high values for the coherence length he would be able to execute an experiment that would decide whether light emission was instantaneous, or not; we will discuss the experiment in the following section (and Einstein's reformulation of Rupp's original experiment in the section after that). Rupp agreed to perform the experiment, and an extensive correspondence between the two men began.[21] Meanwhile, serious criticism of Rupp's Habilitation paper appeared. British spectroscopist Robert Atkinson (1926) claimed that Rupp's results were impossible: because of Doppler effects due to thermal motions of the canal ray atoms, Rupp should not have been able to see interferences for path differences in excess of 3.5 cm in the case of hydrogen. This was the value for a hydrogen source at rest, and there was no reason to believe that the same limitations would be absent in the case of a canal ray source. In fact, due to Doppler effects that were to be expected because of the canal ray's beam motion, the maximum coherence length ought to have been substantially lower than that value, Atkinson argued.

---

[20] Rupp (1926a).
[21] This correspondence, contained in the Einstein Archive at the Hebrew University in Jerusalem, is discussed in van Dongen (2007a), pp. 86-102.

In the end, Rupp could not properly explain his original values, as also Einstein observed in his correspondence with him. The letters exchanged between the two men further reveal that Rupp, after receiving criticism, often changed his data and his explanations for those data. Nevertheless, Einstein showed no reservations when he submitted Rupp's work to the Berlin Academy for publication, and his own theoretical paper stated that Rupp's experiments gave a "full confirmation"[22] of his analysis. Clearly, Einstein must have believed that Rupp was honest when reporting his data, even if confused. Still, he may also have been a bit too eager to see his analysis confirmed. Einstein's physics, in the period of the collaboration with Rupp, was of course removing itself ever more from the practice of experiment, and his concerns would increasingly lie with the highly mathematical unified field theories that would characterize the later part of his oeuvre.[23] He did not assess Rupp's results too critically, perhaps because he was no practitioner of canal ray experiments himself. Yet, he also ignored a fair number of signs that something could have been amiss in Rupp's work; of course, rhetorically, it worked well to propose an experiment, and then to be able to immediately show its successful execution. In any case, Einstein corrected Rupp's numbers quite a few times in the letters that they exchanged, only until Rupp reported precisely what Einstein expected.

Rupp's experimental colleagues did not let him off the hook as easily as Einstein did. In particular a group in Munich, headed by Walther Gerlach and in command of some of the best expertise on the physics of canal rays, doggedly pursued Rupp about his publications. One of their advanced graduate students, Harald Straub, was put to the task of repeating the canal ray experiments, and failed as expected, despite having some of the best facilities in canal ray research at his disposal. Straub could easily explain his failure: the velocity distribution of the atoms in his canal ray beam was too inhomogeneous, leading to a disturbing spread in frequencies, thus inhibiting a stable interference pattern.[24] The implication of Straub's conclusion was that Rupp's work should have been hindered by the same limitations.

Straub's work led to a polemic with Rupp in the very visible *Annalen der Physik*; this hurt Rupp's reputation badly.[25] Other experiments by Rupp were also severely criticized,[26] and by 1934 he saw his academic career and his position at the laboratory of the AEG (*Allgemeine Elektrizitätsgesellschaft*), a leading corporate research lab in Berlin, hang in the balance. Rupp subsequently raised the stakes by publishing impressive work in which he claimed to have artificially produced positrons. Soon, however, his colleagues at the AEG grew suspicious, and Rupp was made to admit that he lacked the accelerator facilities needed to actually carry out such work. The AEG drew up an internal report, condemning Rupp, who was made to retract his most recent publications,[27] after which he was finally dismissed. Rupp suffered a nervous breakdown, and, with the aid of the German Physical Society's embargo, quickly disappeared from the professional literature. He passed away in 1979, having spent the later part of his career in the graphic industry in the GDR.

---

[22] Einstein (1926b), p. 340.
[23] See e.g. van Dongen (2010), in particular Chapter 4, which discusses Einstein's relation to experiment in the context of his unified field theory programme.
[24] Straub (1930).
[25] For a reconstruction of this discussion, see van Dongen (2007a), pp. 102-110.
[26] Overviews can be found in French (1999), pp. 9-15; Franklin (1986), pp. 227-229.
[27] Rupp (1935).

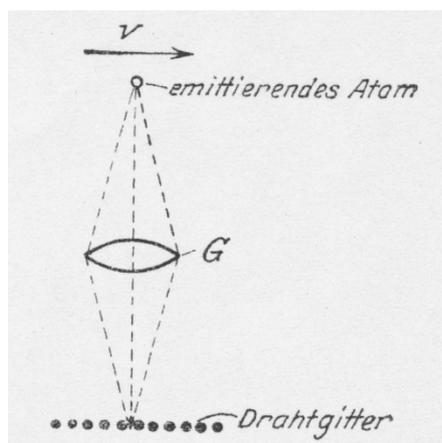

Figure 4. Canal ray atom emitting light while passing behind a wire grid. *Source*: Einstein 1926a.

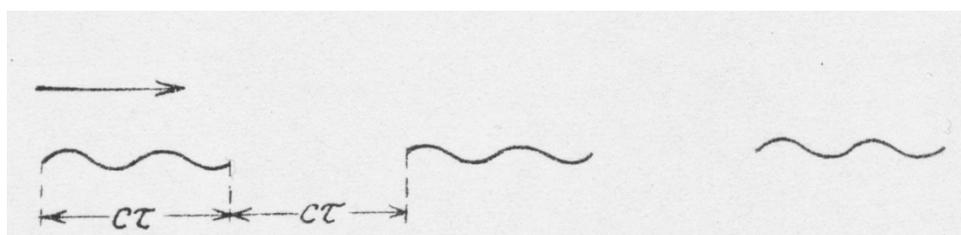

Figure 5. In the classical emission picture, light waves are cut up because of the motion of the canal ray atom behind the grid. *Source*: Einstein 1926a.

**The Wire Grid Experiment**

Let us return to the spring of 1926, when Einstein learnt of the impressive results from Rupp's Habilitation-thesis. He realized that if in Rupp's experiment lens $L_2$ (Figure 3) was replaced by a grid, [28] Rupp had an experiment at hand that should be able to quickly decide whether or not light emission was an instantaneous process; the experiment was soon known in the contemporary literature as the "Wire Grid Experiment".

  If light emission were *not* instantaneous, Einstein believed, the light wave would be cut up as an emitting atom moved behind the openings in the grid (see figure 4; the role of lens $G$ was the same as that of Rupp's lens $L_1$). With the atoms moving at velocity $v$, and the distance between two openings given by $2b$, the wave train would be cut up into pieces that were $c\tau = cb/v$ long, and just as much apart from each other (Figure 5; $\tau$ was the time that it would take an atom to pass behind a slit). In the case that this picture was correct, varying the path difference in the interferometer should produce a varying visibility of the interference pattern. The pattern would completely disappear for values of the path difference of ($n$ is an integer):

(1) $\qquad (2n + 1) \times cb/v.$

---

[28] Lens $L_1$ in figure 2 only produced an unmagnified image of the canal ray beam in slit $B$, and played no role in the interpretations of the experiments—we will omit consideration of this lens in the following.

The interference should have best been visible when the path difference in the interferometer was:

(2) $\qquad 2n \times cb/v.$

Einstein estimated that for reasonable values for the grid and hydrogen canal ray velocity, the partial waves of Figure 4 would be 6 cm long. Rupp claimed to have observed interferences at path differences well above that value, so the variability predicted by (1) and (2) should be easy to observe for him, Einstein stated in a first short publication in *die Naturwissenschaften*.[29]

Yet, he initially expected a different outcome. At the beginning of his correspondence with Rupp and in his short paper just mentioned, Einstein expected that instead of confirming (1) and (2), Rupp would see a stable interference pattern for all values of the path difference. In that case, Einstein thought, the "interference properties of the radiation would have no relation to any periodicity of the emitting atom." In other words: the wave nature of light would not be due to some atomic oscillation, temporally extended. Instead, it would be "conditioned by specific laws of the space-time continuum".[30] As vague as this may sound, such an outcome could open up the possibility that emission actually occurred in an event-like fashion, Einstein likely thought: the light could then be seen to appear on the other side of the grid in an instant, with uninterrupted interference properties, as if only a particle had been emitted and had crossed the openings in the grid. I will return to Einstein's interpretation in a later section.

Einstein changed his mind concerning the outcome of the experiment after rethinking Rupp's original arrangement (Figure 3; with lens $L_2$ instead of a grid). He thought it strange that the Doppler spreading due to the beam motion of the canal ray did not inhibit the formation of an interference pattern in Rupp's *Habilitation* experiment, and subsequently studied its workings in more detail (Atkinson's article had not yet appeared, and Einstein was not yet concerned with Doppler shifts due to thermal motions). He soon found that Rupp's original experiment, too, could decide on the temporal extension of light emission. In fact, since Rupp had seen interferences with his original arrangement, it was clear, Einstein came to believe, that his own newly proposed experiment would fail; that is, that one should expect the classical outcome (variation in the visibility of interference) to be confirmed, as indeed Rupp in the end would do. But for Rupp's original experiment to have decided the issue of instantaneous emission, or even to produce interference at all, something had to have been slightly altered in its arrangement, as we shall see next. Let me end here by pointing out that thinking about beam motion Doppler shifts also led Einstein to produce a fuller account of the experiment outlined above, in which he took

---

[29] Einstein (1926a).
[30] Einstein (1926a).

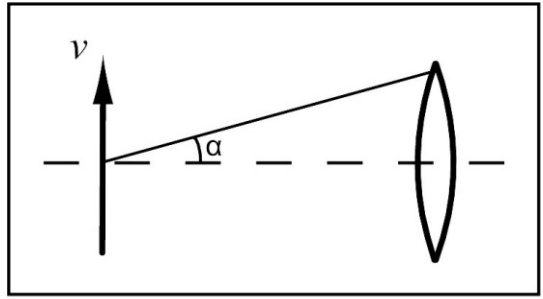

Figure 6. Doppler effect, with the canal ray beam moving up.

diffraction effects into account. This analysis left relations (1) and (2) unaltered however.[31]

**The Rotated Mirror Experiment**

In Rupp's original experiment, Einstein came to realize, wavelengths ($\lambda$) emitted by the canal ray atoms would have to undergo a Doppler shift, due to their motion in the beam (Figure 6):

(3) $$\lambda = \lambda_0 \left(1 - \frac{v}{c}\sin\alpha\right),$$

in which the light makes an angle $\alpha$ with the normal. Einstein feared that the forming of an interference pattern would be obstructed due to these Doppler shifts. Yet, Rupp had claimed to have seen interferences, and Einstein thought of a clever way to explain those: if one were to rotate one of the mirrors in the interferometer, one could tune the path difference such that the rotation would exactly compensate for the Doppler shifts.

Figure 7 helps to explain the need for a mirror rotation to obtain interferences.[32] The canal ray beam's motion makes the upper segment of the light wave coming out of the source slightly blue shifted, and its lower part slightly red shifted. The mirror rotation, as in the diagram, should compensate for this by reducing the path difference in the blue shifted part of the wave, while increasing it for the red shifted part of the wave, exactly such that the ratio between path difference and wavelength is again constant across the wave front. The outgoing signal, after passing through the interferometer's lens (not in the diagram), will hit a point on a screen, and with a stable phase difference contribute to an overall interference pattern. Specifically, rotating one of the mirrors through an angle of $\beta/2$, with $\beta = \frac{v}{c}\frac{d}{f}$ (with $v$ the velocity of the radiating atoms, $c$ the velocity of light, $d/2$ the distance separating the interferometer's mirrors and $f$ the focal length of the lens used), should exactly compensate for the destruction of the interference due to the beam motion Doppler effect.

---
[31] See Einstein (1926b).
[32] For a more detailed technical analysis, see Einstein (1926b); van Dongen (2007a).

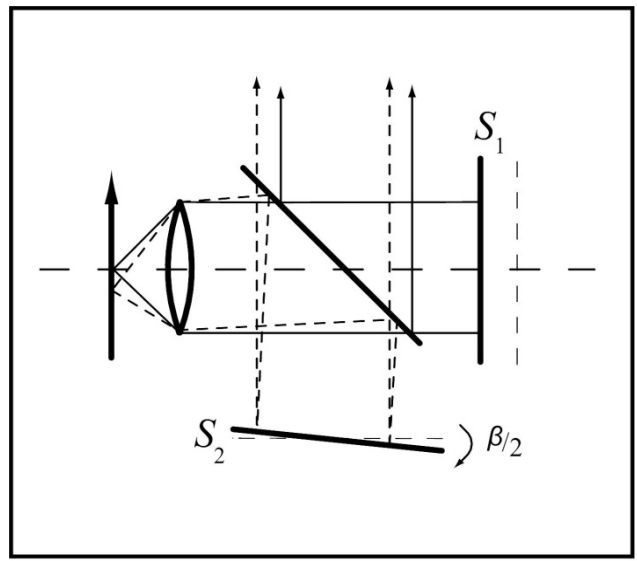

Figure 7. Due to the rotation of mirror $S_2$, the interference pattern should be restored.

As said, Einstein further understood that, with the mirror rotation, Rupp's experiment also showed that light emission was a process that was extended in time; it was because of this realization that his expectations for the outcome of his original Wire Grid Experiment changed. As Figure 7 illustrates, retracing the wave front reflected off the rotated mirror to the source reveals that it was emitted at a different point in the canal ray beam than the wave front that was reflected off the non-rotated mirror with which it interferes. It has also travelled a different distance, and therefore must have been emitted at a different moment. Yet, we see interference, so the interfering wave should have been emitted by an atom that radiates out a coherent signal as it moves up in the beam; light emission, thus, takes an extended lapse of time, Einstein concluded.

Rupp had not mentioned anything about having rotated a mirror in his original arrangement. Yet, since he had seen interferences, Einstein concluded that he—unknowingly—actually must have rotated a mirror. Rupp quickly agreed to this conclusion: an "unconscious rotation of the mirror"[33], as he put it, should explain away destruction of the interference pattern due to the beam motion of the canal ray atoms. However, Doppler shifts due to thermal motions had still not been compensated for, as also Einstein eventually realized, and as Rupp's critics would insist.[34]

As an aside it may be pointed out that this experiment actually assisted in conclusively nailing Rupp down as a fraudster in 1935. Even though Einstein had not made a mistake in his theoretical analysis, he had accidentally drawn the direction of mirror rotation incorrectly in an illustration that he published and had earlier shared with Rupp. The incorrect direction was confirmed in Rupp's final paper,[35] and Rupp even stated that multiple observations backed up this result. In a 1935 publication, Gerlach, together with a co-worker, pointed out the faulty direction to dismiss these claims of Rupp once and for all.[36] Nevertheless, in 1926 Rupp's work was by many taken to lead to the inescapable conclusion that light emission is a process that is extended in time. That is, he claimed to have found a value $\beta$ that was exactly in accordance with Einstein's revised analysis of the experiment. So how did Einstein re-interpret his experiments, after seeing that his initial expectations for instantaneous emission would not be met?

---

[33] Rupp to Einstein, 8 November 1926, Einstein Archive, 20 409.
[34] For more on this, see van Dongen (2007a).
[35] Rupp (1926b).

**Einstein's interpretation**

The first interpretative context that appears relevant is Einstein's idea of a "ghost field". He debated the idea in 1921 with H.A. Lorentz, and their exchange is a good point from which to reconstruct some of Einstein's ideas. H.A. Lorentz summed up Einstein's thinking in a letter to him in 1921, and here Lorentz stated that Einstein believed light consisted of two things: "1. An interference radiation, that occurs according to the normal laws of optics, but still carries no energy. One can for example imagine that this radiation exists in normal electromagnetic waves but with vanishingly small amplitudes … 2. The energy radiation. This consists of indivisible quanta of energy $h\nu$." The 'interference' radiation was emitted along with an individual quantum; an invisible interference pattern of the first actually prescribed a probability distribution for the second, which yielded, upon the radiation of many light quanta, the observed phenomena.[37]

An outcome in line with classical theory might in fact be expected for the Wire Grid Experiment with the above interpretation of light, so this may not literally have been what Einstein initially had in mind in 1926. Unfortunately, it is hard to pin down exactly how he was interpreting the Wire Grid and Rotating Mirror Experiments conceptually: except for a few scattered and incomplete remarks, he remained largely silent on the issue in his papers, both before and after he had changed his expectations. At first Einstein had remarked that if his Wire Grid Experiment would show that "the sine-like character of the wave-field" was not "conditioned by the emitting atom or electron", he suspected that, as quoted, "specific laws of the spacetime continuum" would make light interfere.[38] One would like to think that he had some precise and explicit unified field theory-like intuition for how the "spacetime continuum" would impose its conditions. If he had any, however, he did not share it publicly. For instance, in a lecture in February of 1927 Einstein reportedly only said that Rupp's experiments had shown that radiating particles "do not spit out quanta in a completely irregular fashion", and that one could not ascribe the interference phenomena to "a not yet understood sense for structure directed by space."[39]

Still, Einstein's correspondence makes clear that he had some form of the "ghost field"-interpretation in mind when thinking about the experiments. This is particularly evident after it had become clear that all outcomes were (or, rather, were presumed to be) in line with wave theory predictions. Indeed, Einstein wrote to Rupp that "one must distinguish between the production of the interference field (A) and the energy emission (B). The event-like nature of (B) is certain. Your experiments have proven that (A) is a process that is extended in time."[40] Einstein did not explicitly mention probabilities in his interpretation of the experiments, but some kind of probabilistic conception of the "interference field" seems quite a natural extension on what he wrote to Rupp.

---

[36] Gerlach and Rüchardt (1935).
[37] H.A. Lorentz to Einstein, 13 November 1921, Doc. 298, pp. 347-351 in Kormos Buchwald *et al.* (2009).
[38] Einstein (1926a).
[39] N.N., "Theoretisches und Experimentelles zur Frage der Lichtenstehung", *Zeitschrift für angewandte Chemie*, 40 (1927), 546 (report on lecture by Einstein by unnamed author).

In his elaborate paper in the Academy Proceedings, Einstein motivated his initial particle expectation by pointing out that he thought that the emitted light would be "strictly monochromatic", i.e. would have its frequency narrowly constrained by the energy condition $E = h\nu$. Yet, he informed his readers that a renewed analysis (as we know, this entailed a rethinking of Rupp's original paper) had made him change his mind: he was now convinced that the wave field could not be created instantaneously, and that the wave theory "seems rather to retain its full validity for the creation of the interference field." In a footnote, he added the consideration that we saw paraphrased by Bohr:

> In particular, one may not assume that the quantum process of emission, which in terms of its energy is determined by location, time, direction and energy [cf. particle picture], also has its *geometric* [wave] properties determined by these quantities.[41]

Clearly, this statement raises the question how much of either picture—particle versus wave—could be correct in these experiments. Of course, that was also exactly the question that had prompted Einstein to formulate them to begin with, and it was precisely the point that Bohr, when introducing Heisenberg's new relations to Einstein, referred to and discussed using a reduced version of the Wire Grid Experiment. In any case, the reported outcome of the Einstein-Rupp experiments suggested that when one sets up an optics experiment that can be formulated in terms of waves, one should expect a result that is in line with the wave theory. The expectation of a particle-like instantaneous emission, at least, was concluded to be incorrect.

**The "Experiments of Einstein and Rupp" and the Uncertainty Relations: Bohr's letter**

Let us return to Bohr's letter to Einstein of 13 April 1927. The possibility of avoiding contradictions that Heisenberg's insights afforded seems to have been Bohr's main concern and selling point for the developing quantum theory in his letter. To emphasize the possibilities opened up by the uncertainty relations, Bohr chose Einstein's most recent light quantum test as the most fitting example to introduce them with. At the same time, Bohr's choice also suggests that the Einstein-Rupp experiments were on both his and Heisenberg's mind at this time.

Bohr first discussed a finite wave train, surely relevant for understanding the Wire Grid Experiment. As we saw, he pointed out that it is not strictly monochromatic, and therefore has an "uncertainty" in frequency $\Delta\nu$ and an "uncertainty" in wavelength $\Delta\lambda$. Citing standard results from wave theory, he further made clear that such a wave would take a time $\Delta t = 1/\Delta\nu$ to pass and be at least $\Delta x = \lambda^2/\Delta\lambda$ long. The uncertainty in the description of the waves, "and consequently in the possibility to observe light quanta" stood, he found, "in a peculiar inverse relation to the exactness with which the energy $E = h\nu$ and momentum $p = h/\lambda$ of the quanta are defined." Bohr elaborated this point by stating that "we have $\Delta E \Delta t = h\Delta\nu \cdot \frac{1}{\Delta\nu} = h$ and $\Delta p_x \Delta x = \frac{h\Delta\lambda}{\lambda^2} \cdot \frac{\lambda^2}{\Delta\lambda} = h$." This was "all in agreement with the general relations of simultaneous uncertainty for conjugate variables, which are a direct consequence of the mathematical laws of quantum mechanics according to Heisenberg."[42]

---
[40] Einstein to Emil Rupp, 19 October 1926, Einstein Archive, Hebrew University of Jerusalem, Doc. 70 713.
[41] Einstein (1926b), p. 337.

Heisenberg's new perspective entailed that there would be no contradiction between the wave theory and energy conservation (i.e. strict validity of $E = h\nu$ would be respected—unlike in BKS theory) in the case of Einstein's recent "paradox", Bohr contended. As we have seen, to make this point Bohr collapsed Einstein's Wire Grid Experiment to the problem of a radiating atom moving behind a single slit, which diffracted the emitted light (see Figure 1), and first discussed this from a wave theory point of view. Let us briefly look at this argument again, now that we have seen Einstein's original experiment: the slit limited the time that the atom could be observed, which in turn produced a spread in frequency $\Delta\nu = v/a$. Furthermore, due to diffraction effects, light emitted by an atom at an angle $\alpha = \lambda/a$ to the normal would also reach an observer on the other side of the slit. The Doppler effect that this light would undergo would produce the same relation for the spread in frequency: $\Delta\nu = \nu \cdot \frac{v}{c} \cdot \frac{\lambda}{a} = v/a$. In the particle picture, this would correspond with light quanta with a slightly higher or lower energy—which Bohr in turn accounted for by pointing out that the emitting atoms could transfer various amounts of kinetic energy to the quantum due to back reaction effects. In the end, then, energy conservation would remain valid in individual emissions (and these need not be confined to strictly monochromatic frequencies as Einstein initially expected, we may add). This observation, finally, Bohr saw to be an illustration of Einstein's footnote comment, which he reformulated in his own words: "no possible 'light quantum' description can ever explicitly do justice to the geometrical relations of the 'ray path'." [43]

Bohr's letter thus makes good sense from the perspective of Einstein's experiment. Furthermore, we have learned that Einstein's article, with its suggestive footnote, implied that an experiment that probes the predicted interference, wave-like properties of light, will find these wave properties confirmed. It may therefore not be surprising that Bohr formed and expressed his earliest ideas on complementarity—"according to the description, the different aspects of [Einstein's emission] problem never appear at the same time"—in its context. Still, Einstein was a long way from arguing for anything like a Bohrian concept of complementarity. In fact, his conclusions, certainly in his publication, and even in his private exchanges with Rupp, were analytically clean and modest: he simply limited himself to observe that energy conservation, in particular the light quantum relation, remained exactly valid in emission processes, but this did not preclude that light will exhibit interference properties and can show a spread in frequencies, as followed from the wave theory. Einstein did not put forward some concrete complementarian point of view; yet, the Einstein-Rupp experiments did evidently play on the question of 'when wave; when particle?' at a crucial moment in the history of quantum theory, as we see confirmed through Bohr's letter.

---
[42] Niels Bohr to Albert Einstein, 27 April 1927, pp. 21-24 in Kalckar (1985), on p. 21.
[43] Niels Bohr to Albert Einstein, 27 April 1927, pp. 21-24 in Kalckar (1985), on pp. 22-23.

**Werner Heisenberg and the Einstein-Rupp experiments**

The Einstein-Rupp experiments initially appeared to lay bare a discrepancy between the wave and particle pictures. Heisenberg himself also picked up this element in his discussion of the experiments in his well-known 1929 Chicago lectures, without, however, explicitly referring to his uncertainty relations in his treatment.[44] The experiments' influence on Heisenberg is less obvious than in the cases of Einstein and Bohr. Heisenberg referred to "Einstein's discussions on the relation between wave field and light quanta" as a source of inspiration in his paper on the uncertainty relations, and it is easy to imagine a relation between his presentation of, and discussions with Bohr on the X-ray microscope on the one hand,[45] and Bohr's analysis of the Wire Grid Experiment on the other. Still, Heisenberg did not include an explicit reference to Einstein's paper in his article. The correspondence between Einstein and Heisenberg, at least its portion available in the Einstein Archives, also does not contain any direct discussion of these experiments.[46] Yet, Heisenberg does hint at their influence quite clearly in a lecture he held in 1974 on his "Encounters and Conversations with Albert Einstein". In that lecture, held in Einstein's birth place, Ulm, he recalled meeting and debating Einstein for the first time in "early 1926"—the exact date was April 28, 1926, when Heisenberg gave a seminar in Berlin.[47] This, of course, was precisely when Einstein was exchanging letters almost daily with Rupp on their experiments, and deliberating their outcome.

  Heisenberg would say in 1974 that his private discussion with Einstein on this occasion proved "extraordinarily fruitful" for his subsequent work. In their conversation Einstein had first expressed his novel conviction that it is theory that determines what can be observed if the natural laws are in question and the link between the phenomena and our sensations has become unclear, as in the case of contemporary atomic physics. As Heisenberg recalled, they next turned to what happens in transitions between stationary states: "The electron might suddenly and discontinuously leap from one quantum orbit to the other, emitting a light quantum as it does so, or it might, like a radio transmitter, beam out a wave motion in a continuous fashion". Einstein, in Heisenberg's recollection, pointed to a conflict between a description that can account for the interference phenomena and "the fact of sharp line frequencies."[48] Of course, one immediately recognizes the central question of the Einstein-Rupp experiments.

  Heisenberg had answered Einstein that traditional concepts would not suffice to address the question. This did not convince Einstein: he wanted to know from Heisenberg in what quantum state the continuous emission was then supposed to take place. Heisenberg came up with an analogy with changing film images: in between the projection of one image and another, one sees a hazy blur, and one is unsure which picture is intended. Einstein liked this answer even less, since it suggested that it was "a matter of our knowledge of the atom", yet two people could very well know two different things of the same atom. In the end, Heisenberg concluded, "we separated in the common conviction that a great deal of work still needed to be done."[49] We, in turn, may safely observe that, indeed, the Einstein-Rupp experiments, or at the very least the same problem that motivated them, played an important role in the mix of questions that occupied Heisenberg as he approached the uncertainty relations. It is in any case hard to imagine, given the great impression his conversation with Einstein made on Heisenberg, that Heisenberg did not follow closely the publications discussing the Einstein-Rupp experiments. However, it is not too hard to imagine why, unlike in his 1929 Chicago lectures, he did not mention them explicitly in 1974.

---

[44] Heisenberg (1930, p. 60), gave the same discussion of the Wire Grid Experiment as Bohr in his letter, yet slightly altered the logic of the latter's argument: he too limited the problem to that of a radiating atom behind a slit, but credited Bohr with the insight that accounting for diffraction and the Doppler effect solved the supposed discrepancy, without mentioning back reaction effects nor indeed the uncertainty relations.

[45] See Heisenberg (1927), pp. 174-175; 197-198.

[46] See Belousek (1996) for the exchange between Heisenberg and Einstein on a retracted hidden variable theory that Einstein proposed soon after Bohr's letter, in May of 1927.

[47] See Cassidy (1991), p. 235.

[48] As in Heisenberg (1989), p. 114.

**Uncertainty in the Einstein-Rupp experiments after Emil Rupp's demise**

The most natural discussion of the Einstein-Rupp experiments in the context of the uncertainty relations was given by H. Billing and appeared in 1938.[50] Billing was another graduate student in Gerlach's laboratory in Munich who had redone the Rotated Mirror Experiment (his analysis also focused on this experiment, and not the Wire Grid Experiment). By 1938, apparatuses to produce homogeneous canal rays had improved substantially and Billing, unlike Straub, did succeed in confirming Einstein's analysis of the experiment—still, his coherence lengths were no less than three orders of magnitude smaller than those of Rupp (Rupp's name was quite markedly absent from the article).

Billing wrote a concluding section for his experimental paper in which he explained that his result for the Rotated Mirror Experiment was not in opposition to the "photon concept, as it has been altered by the Heisenberg uncertainty relation."[51] He had used a Fabry-Perot interferometer (producing the same fringes as a Michelson interferometer or plane parallel dielectric slab), and reminded his readers that this, for perpendicularly in-falling light, has a resolving power of:

(7) $$\frac{\Delta \nu}{\nu} = \frac{\Delta \lambda}{\lambda} = \frac{\lambda}{d}.$$

The frequency, thus, had an "uncertainty" of $\Delta \nu = \nu \lambda / d = c/d$, and the uncertainty in the energy was given by $\Delta E = h \Delta \nu = hc/d$. The uncertainty relation for energy and time, $\Delta E \Delta t = h$, then implied that in the Rotated Mirror Experiment, the exact time of emission could only be determined up to:

(8) $$\Delta t = \frac{d}{c}.$$

As the emitting atom moves with velocity $v$, the exact location of emission in the canal ray beam could only be found with an inaccuracy of:

---

[49] Heisenberg (1989), p. 115.
[50] Billing (1938).
[51] Billing (1938), p. 591.

$$(9) \qquad \Delta s = \frac{vd}{c}.$$

      Billing had earlier argued that, in a wave picture, the two points in the canal ray beam from which the interfering light came (see Figure 7) were the same distance $v \cdot d/c$ apart: the optical path difference between the interfering wave fronts was $d$; travelling the extra distance would take a time $d/c$, during which the emitting atom would move $v \cdot d/c$ up the beam. Thus, the uncertainty relation entailed that the location of emission of a light quantum cannot be established with any greater accuracy than the distance along which the atom had radiated according to the wave theory. Billing concluded that, due to the uncertainty relations, the statement that one has brought light from the two emitting points in the beam in Figure 7 to interfere is "pointless."[52] In any case, Billing's account does make clear once more, as Bohr's letter already suggested, that the Einstein-Rupp experiments are an excellent way to illustrate the uncertainty relations.

**Conclusion: fraud and physicist histories**

As the above has hopefully shown, the Einstein-Rupp experiments played a direct and crucial role in the history of quantum mechanics: they asked a central question at a key moment in time, and the historical actors recognized them for it. They involved them in their attempts to get an ever stronger hold on the theory of the quantum. Yet, the role of the Einstein-Rupp experiments—in particular Rupp's part—has not usually been made explicit when this period is discussed. In fact, it even appears as if they have been deliberately omitted from the historical record.

      For instance, we saw that the letter that Bohr wrote to Einstein is included in volume 6 of Bohr's *Collected Works*, edited by physicist Jørgen Kalckar. Kalckar does provide the reader with a reference to Einstein's paper in his annotation, but states that its conclusions derived from "general arguments"—no mention of Emil Rupp and his experiment is made.[53] An even stronger example of apparently willful neglect is found in Abraham Pais' work. In his biography of Bohr, Pais discussed what could have "stimulated [Max] Born's radical new ideas" of interpreting the wave function probabilistically in 1926. According to Pais, Born's "inspiration came from Einstein."[54] He next makes clear that he meant Einstein's ghost field ideas for light quanta, and informs us that Einstein never published those. Pais omits to mention, however, that these ideas were very much the focus of attention due to the Einstein-Rupp experiments exactly when Born produced his interpretation.

      The omission is even more problematic in Pais' biography of Einstein. In his '*Subtle is the Lord…*',[55] Pais gives an impressive overview of Einstein's scientific work, and places it in the development of contemporary physics. Even if the book may be called Whiggish, the physics is rich and insightful, and Pais pays considerable attention to less well-known aspects of Einstein's oeuvre, such as several details of his unified field theory work or an earlier, failed canal ray experiment from 1921. He also included a twenty page appendix that supplies short biographies of Einstein's

---

[52] Billing (1938), p. 592.
[53] Kalckar (1985), see its introduction for an English translation of Bohr's letter (pp. 21-24), and pp. 418-421 for the original German version. The footnote mentioned is found on p. 22.
[54] Pais (1991), pp. 287-288 on p. 287; see also his earlier article in *Science* on this subject, Pais (1982b).
[55] Pais (1982a).

collaborators—he did not include Rupp in that appendix, however. In fact, he did not even mention once the Einstein-Rupp experiments, nor Emil Rupp, in his nine chapters long discussion of Einstein's involvement with the quantum theory, even though Einstein published two very substantial papers on these experiments in the key year of 1926, and even though, as we have seen here, these experiments played a direct and relevant role in early quantum debates. Clearly, Pais chose to edit Rupp out of Einstein's life and remove him from the history of quantum mechanics.[56]

The question then is: why? Unfortunately, Abraham Pais is no longer with us, so we can only guess. Still, deliberating on the question is legitimate, as it may inform us about biases and blind spots in writing the history of physics. When contemplating these omissions, I am first reminded of Thomas Kuhn's recollection of his 1962 interview with Niels Bohr on the latter's quantum atom. Kuhn asked Bohr about his earlier attempts—incorrect from the perspective of the 1913 Bohr model—to quantize Rutherford's atom without knowledge of the Balmer series. Bohr denied any such attempt, but Kuhn knew this to be wrong on the basis of documentary evidence and certain passages in Bohr's original paper, which he pointed out to Bohr. Bohr still remained silent on the episode. Kuhn ascribed Bohr's reaction to physicists' tendency to review the past in light of the current state of affairs. This then leads to willful or accidental (as apparently in Bohr's case) distortions of memory, and produces "the linearized or cumulative histories familiar from science textbooks."[57] A similar mechanism may have been involved in the neglect of the Einstein-Rupp story. It is not even difficult to come up with a rationalization for it: as Rupp had committed fraud, his experiments never *actually* established facts about nature, so they should or could not *in fact* have influenced events, and therefore can be omitted. Such an implicit rationalization would be aided by the circumstance that Einstein's final predictions turned out to be in line with the result that the developed theory of quantum mechanics would propose. Yet, the experiments *did* of course impact events: for one, Rupp's work made Einstein change his mind, at a crucial moment, about the outcome of the Wire Grid Experiment and the instantaneous nature of light emission. Secondly, we have seen how they helped Bohr to shape his views on complementarity and the uncertainty relations. Of course, the omission of the important role of these experiments distorts the historical record and denies us the possibility to get a full understanding of how science works.

Rupp distorted the facts, and so have a number of historians. Rupp committed his sin in an attempt to advance and maintain his professional career, but what interest or values were served by the omissions of the latter? These omissions seem accidental in some cases, yet fairly blatant in others. Accidental omissions indicate that perhaps not enough research was done that begins with a systematic and critical reviewing of primary sources or even just bibliographies. Deliberate omissions, and their repetition, indicate that some historians may tacitly have observed a shared interest in leaving out Rupp's role.

One does occasionally come across Rupp in older historiography. Usually, however, his place is on the margins and he is judged to be a black sheep. For instance, H.B.G. Casimir mentioned Rupp in his autobiography when he illustrated how his own mentor, Paul Ehrenfest, dismissed pomp: Rupp had called his experiment of colliding electrons 'an electric analogue of the Compton effect', upon which Ehrenfest quipped that shooting a tail of a bird could be called the biological analogue of the photo-electric effect.[58] The anecdote suggests that Rupp's physics was of only secondary importance, and a little silly (even if Casimir also called his early work "meritorious"). It further illustrates another aspect of the Rupp case: it was part of the living memory of Casimir and Pais's generation of elite physicists. That generation also held up Albert Einstein as an icon of science, and, perhaps even more importantly, saw quantum theory and its Copenhagen interpretation as singularly important achievements for which proselytizing still had to be done.[59] Clearly, tainting either by attributing a prominent role to the Einstein-Rupp experiments would go against larger shared values and interests, to which the value of historical accuracy had to yield.

---

[56] I have been told that Pais had systematically compiled binders that contained copies of every single journal paper by Einstein, together with his own notes on them (A.J. Kox, private communication). If he had used one of the standard Einstein bibliographies, such as the one contained in the Schilpp (1949/1997) volume, or the one put together by Boni et al. (1960), it should have been inevitable that he would have come across references to the articles on the Einstein-Rupp experiments.
[57] Kuhn (1984), pp. 247-249.

Modern science has always been an intrinsically moral enterprise[60] and observing and hence allowing something as abhorrent as fraud a substantial, proximate role in reports that focus on its patron saints has proven problematic to scientist-biographers in Rupp's case. Neat black-and-white accounts better serve historiography that aims at establishing and portraying model science; hence Rupp is omitted or only awarded a safe place in the margins. (Professional historians of science—obviously the distinction between the two groups is quite inexact—that have overlooked Rupp should perhaps rather be reproached for not engaging with their sources closely enough.) Fraud, as one of the most severe examples of scientific misconduct, clearly violates the ideal self-image of the scientist and the code of conduct he gets inculcated through, e.g., textbooks. Therefore, the scientific community is still often greatly surprised by the exposure of its occurrence, and seems to underestimate its scale, just as the scale of other professional misconduct.[61] Yet, the actual practice of science and the life in a laboratory exhibit many more shades of grey than clear black-and-whites—as we know at least since the science studies of the 1970s and 1980s. In this sense, science does not depart from other human activity to which moral judgement is wholly intrinsic.

The black cases are uncomfortable reminders of the more prevalent shades of grey, which is why the kneejerk reaction is to isolate and marginalize their role; for, in the case of science, they ultimately undermine the public authority of the scientist. These are however not issues to elaborate on further here, except for observing that writing history, too, is inherently and obviously a moral enterprise, in which epistemic virtues—values and practices internalized and agreed by a group of scholars as inducing knowledge—play a key role.[62] Scientist-historians that came across Rupp, facing a typical conflict of conscience due to their crossing of disciplinary lines, weighed the values of the historiographer to the interests of the scientist, and chose to let the second prevail.

---

[58] Casimir (1983), p. 86.
[59] On enforcement of the Copenhagen hegemony see e.g. Freire (2005, 2009).
[60] See for example Daston (1995); Shapin (2008); see also Daston and Galison (2007) on 'epistemic virtues'.
[61] These points are illustrated by the recent high-profile fraud case of Dutch social psychologist Diederik Stapel, even if in the case of his field there had already been a longer held concern about misconduct; see e.g. "Fraud case seen as a red flag for psychology research", *New York Times*, 2 November 2011; Abma (2013).
[62] On the parallels between the roles of epistemic virtues in science and historiography, see Paul (2011).

Einstein committed the same sin, although it was not committed in an attempt at writing history. In 1936, renewed interest in Germany arose about the execution of the Einstein-Rupp experiments: improvements in canal ray beam production promised that the experiments could by then really be done. Max von Laue informed Einstein, who was by now in the USA, about these developments, and debated the experiments' possible outcome with him. Einstein grew a little impatient with von Laue's reasonings, and stated that the latter had "not appreciated the point that makes my considerations of those days meaningful." The experiments, Einstein now felt, had been formulated as cases for which "our knowledge would make a decision possible, even without carrying out an experiment." In his exchanges with von Laue, Einstein did not once mention Rupp, and stated that "of course, also back then they did not require any confirmation by experiment."[63] Reading between the lines, we see Einstein refashion his recollections to reduce the episode with Rupp to an exercise in only theoretical physics, in which some unnamed experiment had played a marginal, perhaps rhetorical, but certainly entirely superfluous role. As it would turn out, the experiments were dropped from physicists' historiography and collective memory altogether.

---

[63] Einstein to Max von Laue, 29 August 1936, Einstein Archive, entry 16 113.